\documentclass[12pt,final]{article}
\usepackage{graphicx}
\usepackage{epsfig}
\usepackage{dcolumn}
\usepackage{bm}

\usepackage{amssymb,amsmath}
\usepackage{gensymb}
\usepackage{multirow}
\usepackage{booktabs}
\makeatletter
\def\@biblabel#1{(#1)}
\makeatother
\usepackage{setspace}

\usepackage{booktabs,rotating}
\usepackage{fancyhdr}
\usepackage{booktabs,caption}
\usepackage[flushleft]{threeparttable}
\usepackage{enumerate,subfigure,tabularx,longtable}
\usepackage{longtable}
\usepackage[utf8x]{inputenc}
\UseRawInputEncoding
\usepackage{gensymb}
\usepackage {threeparttable} 

\newcommand{\et}{\textit{et al.}}

\newcommand{\comment}[1]{}

\usepackage{lineno}   

\def\gsim {\mbox{\hbox{ \lower-.6ex\hbox{$>$}
\kern-1.12em \lower.5ex\hbox{$\sim$}\kern+.35em}}}
\def\lsim {\mbox{\hbox{ \lower-.6ex\hbox{$<$}
\kern-1.12em \lower.5ex\hbox{$\sim$}\kern+.35em}}}
\makeatletter
\let\@fnsymbol\@arabic
\makeatother

\begin{document}


\title{\vspace{-2.0cm}
	\Large \textbf{Molecular Insights into Gas Nanofilms Confined Between Bulk Liquid Phases}}

\author{Yafan Yang$^{\dag,\ddag,\ref{fn:1}*}$, Zufeng Zuo$^{\dag,*}$, Xingyu Zhao$^{\ddag}$,\\ Shuyu Sun$^\S$, and Denvid Lau$^{\ddag,\ref{fn:1}*}$ \\
	\\[-15pt] 
	\small  $^{\dag}$State Key Laboratory of Intelligent Construction and Healthy Operation \\
	\small  and Maintenance of Deep Underground Engineering, \\
	\small  China University of Mining and Technology, 
	\small  Xuzhou, Jiangsu, China. \\
	\small $^\ddag$Department of Architecture and Civil Engineering, \\
	\small  City University of Hong Kong, 
	\small  Hong Kong, China. \\
	\small $^\S$School of Mathematical Sciences, Tongji University, Shanghai, China.\\
}

\date{\today}
\maketitle

\footnotetext{\label{fn:1}$^*$ To whom correspondence should be addressed, e-mails: yafan.yang@cumt.edu.cn; zufeng\_zuo@163.com; denvid.lau@cityu.edu.hk.}

\newpage
\begin{abstract}

Nanometer-thick fluid films play a critical role in confined multiphase processes, yet the thermodynamics and stability of free gas nanofilms remain poorly understood compared with their liquid counterparts. Here, molecular dynamics (MD) simulations are employed to systematically investigate gas nanofilms confined between bulk liquid phases using Lennard-Jones argon as a model system. 
%
The results show that the surface tension decreases exponentially with decreasing film thickness, accompanied by an increasing magnitude of the negative disjoining pressure.
Upon thinning, the planar gas film undergoes a distinct morphological transition from a stable planar state to a transient or persistent spherical bubble through the formation and growth of a liquid bridge. 
The film surface area strongly affects its thermodynamic properties, with larger areas producing stronger thickness dependence and larger deviations from classical density functional theory (cDFT) predictions. The closer agreement between MD and cDFT at smaller surface areas suggests that the discrepancy primarily arises from thermal capillary-wave fluctuations, which are included in MD but omitted in mean-field cDFT.
Moreover, at small film thicknesses, the magnitude of the disjoining pressure increases with decreasing temperature, consistent with the enhanced sensitivity of the confined gas phase to thickness variations and contrasting with the trend generally reported for liquid nanofilms.
These findings provide molecular insights into the thermodynamics and stability of gas nanofilms, with implications for confined multiphase transport and droplet coalescence.

\end{abstract}

\clearpage

\newpage
\section{Introduction}


The transition toward sustainable energy technologies has driven rapid advances in subsurface energy systems, including enhanced oil recovery, geological carbon dioxide sequestration, underground hydrogen storage, and geothermal energy exploitation~\cite{ding2021pore,iglauer2017co2,ali2025recent,jolie2021geological}. A fundamental characteristic shared by these applications is the multiphase transport and displacement of fluids within complex porous geological formations, where interfacial phenomena critically control wettability, capillary pressure, phase connectivity, and fluid mobility. 
In particular, unconventional and tight geological formations often contain nanoporous structures with pore sizes ranging from a few to tens of nanometers, where fluid behavior is strongly influenced by confinement effects~\cite{salahshoor2018review,li2017thickness}. 
As pore dimensions approach the molecular scale, surface forces and molecular interactions are comparable to or even exceed bulk thermodynamic contributions, resulting in interfacial behaviors that deviate significantly from classical continuum descriptions. 

Among the various interfacial processes relevant to subsurface energy applications, nanometer-thick fluid films have attracted growing attention due to their critical roles in regulating phase behavior and transport at confined interfaces~\cite{israelachvili2011intermolecular,firoozabadi2016thermodynamics}. Such nanofilms can form transiently between approaching fluid interfaces or persist at fluid-solid boundaries, governing processes including droplet and bubble coalescence, wettability alteration, interfacial adhesion, and molecular transport. Their formation, stability, and rupture directly influence fluid displacement efficiency, phase connectivity, and capillary trapping in porous media. Nevertheless, despite their technological importance, the fundamental mechanisms controlling the behavior of fluid nanofilms remain incompletely understood, largely because experimentally resolving the molecular structure and thermodynamic properties of these ultrathin confined interfaces remains highly challenging~\cite{bhatt2002molecular,stephan2020enrichment}.

Extensive molecular simulation studies have been conducted to characterize the key thermodynamic and mechanical properties of liquid nanofilms, particularly the disjoining pressure and its dependence on film thickness. Bhatt et al.~\cite{bhatt2002molecular} investigated the disjoining-pressure isotherms of free liquid nanofilms composed of a pure Lennard-Jones (LJ) fluid. By applying the Gibbs-Duhem relation, they obtained the disjoining pressure from the variation of chemical potential with film thickness and reported negative disjoining pressures for the pure LJ system. Subsequently, Peng et al.~\cite{peng2015methodology} proposed an alternative approach based on interfacial tension calculations to determine the disjoining pressure. 
However, significant discrepancies were observed between the disjoining-pressure profiles obtained from these two approaches. In our recent studies, we showed that these discrepancies can be substantially reduced by incorporating long-range dispersion interactions, adopting a thermodynamically consistent definition of film thickness, and accounting for variations in bulk liquid density with film thickness~\cite{yang2026resolving,jianzhou2026estimating}. Beyond simple LJ fluids, molecular simulations have been extended to more chemically realistic systems, including multicomponent liquid nanofilms~\cite{bhatt2003molecular,bhatt2004monte,benet2014disjoining,peng2015surface,peng2016accelerated} and liquid films adsorbed on solid substrates~\cite{fang2019structure,sun2020molecular,li2024hygroscopic,han2008disjoining}, providing insights into the effects of molecular composition and solid-fluid interactions on nanoscale film stability and interfacial properties.

Moreover, previous studies have examined the stability of free liquid nanofilms and the effects of film thickness, surface area, and temperature on their surface properties. Filippini \et~\cite{filippini2014communication} investigated the rupture of liquid nanofilms and argued that the apparent decrease in surface tension with decreasing film thickness was primarily associated with the formation of holes, rather than an intrinsic thickness dependence, provided that the film remained intact. In contrast, Peng \et~\cite{peng2016quantitative} demonstrated that the surface tension of stable aqueous nanofilms can exhibit a genuine thickness dependence, with distinct thickness regimes associated with different stability characteristics and disjoining-pressure behaviors. They further showed that lateral system size introduces a finite‑size effect on the calculated surface tension; this effect is appreciable at relatively small lateral dimensions but becomes negligible when the lateral size exceeds approximately 4 nm. Subsequent studies have also demonstrated a pronounced temperature dependence of the disjoining pressure in liquid nanofilms, with higher temperatures generally leading to lower disjoining pressures in the thin-film regime~\cite{bhatt2002molecular,peng2015methodology,yang2026resolving,jianzhou2026estimating,peng2016quantitative}. Collectively, these studies demonstrate that film thickness, surface area, and temperature can substantially affect the thermodynamic and mechanical properties of liquid nanofilms, while highlighting the important roles of confinement and thermal fluctuations in governing their surface behavior.

Despite these advances in the study of free liquid nanofilms, investigations of free gas nanofilms remain scarce, leaving their fundamental thermodynamic and structural behavior poorly understood. Existing studies on free gas films have mainly focused on the dynamic behavior at larger scales, particularly their role in mediating droplet coalescence~\cite{chan2011film,dai2008mechanism,janssen2006axisymmetric}. 
To date, preliminary understanding of gas nanofilms is based on classical density functional theory (cDFT)~\cite{yang2026accounting}. However, as a mean-field approximation, cDFT does not explicitly account for thermal capillary-wave fluctuations, which have been shown to influence surface roughness and mechanical stability in free liquid nanofilms\cite{yang2026accounting,macdowell2014disjoining}. Moreover, it remains unclear whether the thickness-, area-, and temperature-dependent behaviors established for liquid nanofilms can be directly extended to gas nanofilms, given the fundamentally different state of the confined phase. Molecular dynamics (MD) simulations naturally capture thermal molecular fluctuations and provide atomistic resolution of the evolving gas-liquid surfaces; however, systematic MD studies of free gas nanofilms have not yet been reported. This lack of molecular-level understanding represents a key knowledge gap in nanoscale thin-film thermodynamics.

In this work, we employ MD simulations to investigate the thermodynamics and stability of gas nanofilms confined between bulk liquid phases. Unlike continuum-based approaches, the present MD study explicitly accounts for molecular fluctuations and surface rearrangements, enabling a direct characterization of the structural evolution and thermodynamic properties of gas nanofilms. A thermodynamic framework based on the Gibbs surface formulation is established to consistently relate the film surface tension to the disjoining pressure. Using argon as a model LJ system, we systematically investigate the effects of film thickness, lateral system size, and temperature on the evolution and stability of confined gas nanofilms. These results provide molecular-level insights into the formation, stabilization, and transition mechanisms of gas nanofilms, extending the current understanding of nanoscale surface phenomena beyond the relatively well-studied regime of liquid thin films.

\section{Method}

\subsection{Gas nanofilm thermodynamics}

\begin{figure}[tb]
	\begin{centering}
		\includegraphics[width=0.9\textwidth]{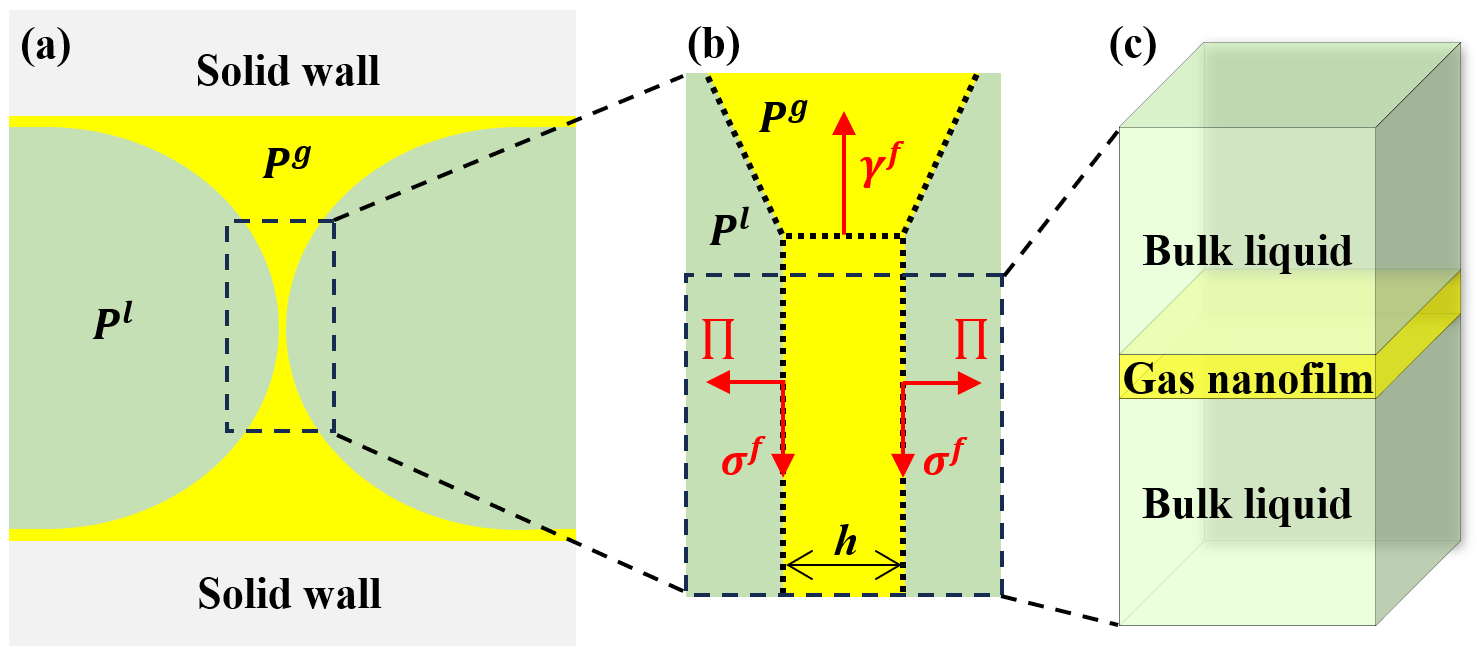}
		\caption{(a) Schematic illustration of a gas nanofilm ($f$) confined within a pore. The nanofilm is connected to the bulk gas ($g$) through a Plateau-border meniscus and is sandwiched between two bulk liquid ($l$) phases. (b) Mechanical equilibrium of the confined nanofilm, illustrating the disjoining pressure $\Pi$, film tension $\gamma^f$, and surface tension $\sigma^f$ acting on the film. (c) Molecular dynamics simulation setup employed to determine the film surface tension $\sigma^f$.}
		\label{fig:z1}
	\end{centering}
\end{figure}

To establish a thermodynamic framework for a single-component gas nanofilm confined between two bulk liquid phases (see Fig.~\ref{fig:z1}), we apply the Gibbs surface formulation following the derivation described by Toshev and Ivanov for thin liquid films~\cite{toshev1975thermodynamics,ivanov1975thermodynamics2}. Consider a plane-parallel thin gas film of cross-sectional area $A$ and thickness $h$, corresponding to a film volume of $V^f = A h$. The film is surrounded by its bulk liquid phase ($l$) and contiguous bulk gas phase ($g$).

Following Gibbs' treatment, any extensive thermodynamic property $E$ can be expressed as an excess quantity $\tilde{E}$ by subtracting the homogeneous bulk phase contributions:
\begin{equation}
	\label{eq:excess}
	\tilde{E} = E - E^l - E^g.
\end{equation}

The fundamental differential equations for the total Helmholtz free energy $F$ of the film system ($i.e.$, Fig.~\ref{fig:z1}c) and its corresponding bulk reference phases are given by:
\begin{align}
	dF &= -S dT - P^l dV + \gamma^f dA + \mu dN, \label{eq:dF_real} \\
	dF^g &= -S^g dT - P^g dV^f + \mu dN^g, \label{eq:dF_gas} \\
	dF^l &= -S^l dT - P^l dV^l + \mu dN^l, \label{eq:dF_liq}
\end{align}
where $S$ and $T$ denote the entropy and temperature, respectively; $P$ is the pressure; $\gamma^f$ represents the film tension ($i.e.$, the total lateral force per unit length acting on the film perimeter); $\mu$ is the chemical potential; and $N$ is the number of moles of the single component.

By subtracting Eqs.~\eqref{eq:dF_gas} and \eqref{eq:dF_liq} from Eq.~\eqref{eq:dF_real}, and applying $dV = dV^f + dV^g$ together with the mechanical definition of the disjoining pressure $\Pi = P^l - P^g$, we find that the differential excess Helmholtz free energy $d\tilde{F} = dF - dF^g - dF^l$ reduces to:
\begin{equation}
	d\tilde{F} = -\tilde{S} dT - \Pi dV^f + \gamma^f dA + \mu d\tilde{N},
	\label{eq:dF_excess}
\end{equation}
where $\tilde{S} = S - S^l - S^g$ and $\tilde{N} = N - N^l - N^g$ represent the excess entropy and excess molar quantity according to Eq. \eqref{eq:excess}, respectively.

Since $\tilde{F}$ is a first-degree homogeneous function with respect to its extensive parameters ($V^f$, $A$, and $\tilde{N}$), integration via Euler's theorem yields:
\begin{equation}
	\tilde{F} = -\Pi V^f + \gamma^f A + \mu \tilde{N}.
	\label{eq:F_excess_integrated}
\end{equation}

Rusanov related the film tension $\gamma^f$ to the film surface tension $\sigma^f$ with the following equation~\cite{toshev1975thermodynamics}:
\begin{equation}
	\gamma^f =  2\sigma^f + \Pi h.
	\label{eq:sigma_f_def}
\end{equation}

Here, $\sigma^f$ denotes the surface tension of a single surface of the film. This equation states that the total boundary force $\gamma^f$ comprises both the dual-surface tension $2\sigma^f$ and a volumetric force component $\Pi h$ resulting from the pressure differential across the film.

Substituting Eq.~\eqref{eq:sigma_f_def} into Eqs.~\eqref{eq:dF_excess} and~\eqref{eq:F_excess_integrated} allows the formulation of free energy in terms of film surface tension $\sigma^f$:
\begin{align}
	d\tilde{F} &= -\tilde{S} dT - \Pi A dh + 2\sigma^f dA + \mu d\tilde{N}, \label{eq:dF_sigma_f} \\
	\tilde{F} &= 2\sigma^f A + \mu \tilde{N}. \label{eq:F_sigma_F_integrated}
\end{align}

Substituting Eq. \eqref{eq:F_sigma_F_integrated} into Eq. \eqref{eq:dF_sigma_f} yields the final expression for the fundamental differential change in film surface tension:
\begin{equation}
	2 A d\sigma^f = -\tilde{S} dT - \Pi A dh - \tilde{N} d\mu.
	\label{eq:d_sigma_f_final}
\end{equation}

The Gibbs' definition of the dividing surface gives the thickness of the gas film:
\begin{equation}
	h=\frac{N/A-\rho_{l}L}{\rho_{g}-\rho_{l}},
	\label{eq:film_thickness}
\end{equation}
where $L$ is the length of the film system in the direction perpendicular to the surface, and $\rho$ denotes the bulk density.

Adopting this definition of $h$ for which the excess molar amount vanishes ($\tilde{N} = 0$), Eq. \eqref{eq:d_sigma_f_final} under isothermal equilibrium ($dT = 0$) reduces to:
\begin{equation}
	2d\sigma^f = -\Pi dh.
\end{equation}

Finally, the disjoining pressure $\Pi$ of the gas nanofilms can be evaluated as:
\begin{equation}
	\Pi = -2 \left( \frac{\partial \sigma^f}{\partial h} \right)_T
	\label{eq:disjoining_pressure}
\end{equation}

This expression shares the same form as that derived for liquid  nanofilms~\cite{ivanov1975thermodynamics2}, albeit with a different formula for the film thickness $h$. Note that this formulation can be extended to multicomponent systems~\cite{ivanov1975thermodynamics2}. However, for such systems, the contributions from the surface excess quantities and the chemical potentials of all components should be explicitly incorporated into Eq.~\eqref{eq:d_sigma_f_final}. Accordingly, Eq.~\eqref{eq:disjoining_pressure} is only valid when the chemical potentials of the other species remain constant during the variation of film thickness. 

\subsection{Simulation details}
All MD simulations were performed using the LAMMPS package~\cite{thompson2022lammps}. Argon was modeled using a reparameterized LJ force field~\cite{mendez2022argon,barker1971liquid}. As demonstrated in our previous study, neglecting long-range dispersion interactions can lead to significant errors in the calculated $\Pi$~\cite{yang2026resolving}. Accordingly, short-range LJ interactions were truncated at a cutoff distance equal to half of the shortest simulation box length, whereas the long-range dispersion interactions were evaluated using the particle-particle particle-mesh (PPPM) method with a relative force accuracy of $10^{-4}$.

To maintain the stability of the confined gas nanofilm, the simulation cell was designed with a finite lateral area to suppress long-wavelength capillary fluctuations that could otherwise induce film rupture~\cite{bhatt2002molecular}. The influence of the lateral box size on the interfacial properties was investigated in this work. The box length normal to the fluid surface was fixed at 140.16~\AA{}. Periodic boundary conditions and the minimum image convention were applied in all three spatial directions.

Simulations were carried out in the canonical ($NVT$) ensemble with temperature controlled by a Nos\'e-Hoover thermostat. The equations of motion were integrated using the velocity-Verlet algorithm with a time step of 5~fs. Each simulation consisted of a 7.5~ns equilibration stage followed by a 22.5~ns production stage. The production trajectory was divided into three equal blocks to estimate the statistical uncertainty.

The film surface tension was obtained from the MD simulations with a system setup shown in Fig.~\ref{fig:z1}c using the Bakker equation~\cite{bakker1928kapillaritat,green1960molecular}:
\begin{equation}
	\label{eq:6}
	\sigma^{f}=\frac{1}{2}\int_{-\infty}^{+\infty}
	\left[
	P_{zz}-\frac{P_{xx}+P_{yy}}{2}
	\right]\mathrm{d}z,
\end{equation}
where $P_{xx}$, $P_{yy}$, and $P_{zz}$ are the diagonal components of the pressure tensor along the $x$-, $y$-, and $z$-directions, respectively, with the $z$-direction normal to the fluid surface. 

The disjoining pressure was determined from the MD-derived surface tension as a function of film thickness. The surface tension data were first fitted using an exponential function \cite{yang2026resolving,li2024hygroscopic,jianzhou2026estimating}:
\begin{equation}
	\label{eq:888}
	\sigma^f=a \cdot e^{b\cdot h}+c,
\end{equation}
where $a$, $b$, and $c$ are fitting parameters with units of $mN/m$, $\textup{\AA}^{-1}$, and $mN/m$, respectively. The disjoining pressure was subsequently obtained by substituting Eq.~\eqref{eq:888} into Eq.~\eqref{eq:disjoining_pressure}, yielding:
\begin{equation}
	\label{eq:8882}
	\Pi=-20\cdot  a\cdot b \cdot e^{b\cdot h},
\end{equation}
where $\Pi$ is in units of $MPa$.

When determining the film thickness using Eq.~\eqref{eq:film_thickness}, the bulk liquid density was obtained by averaging the density profile over the central region of the liquid phase, where a plateau density was observed. Since no bulk gas reservoir was included in the simulation cell, the bulk gas density was approximated from the average density in the central region of the thickest gas nanofilm with the same surface area and temperature.

It should be noted that, in our previous study of liquid nanofilms, we demonstrated that approximating the bulk density in the calculation of $h$ can introduce significant errors in the estimated thickness of water nanofilms and moderate errors for argon nanofilms~\cite{jianzhou2026estimating}. In the present work, however, this approximation is expected to have a negligible effect. This is primarily because, under the thermodynamic conditions considered, the bulk gas density is one to two orders of magnitude lower than the bulk liquid density. Furthermore, the disjoining pressure is relatively small and is governed primarily by variations in the liquid pressure, $P^l$, rather than the gas pressure, $P^g$, according to the chemical-potential-pressure relationship for argon~\cite{bhatt2002molecular}. Consequently, even relatively large uncertainties in the estimated bulk gas density have a negligible influence on the calculated film thickness.

\section{Results and Discussion}

\subsection{Effect of Film Thickness}

\subsubsection{Surface property isotherms}
\begin{figure}[tb]
	\begin{centering}
		\includegraphics[width=0.5\textwidth]{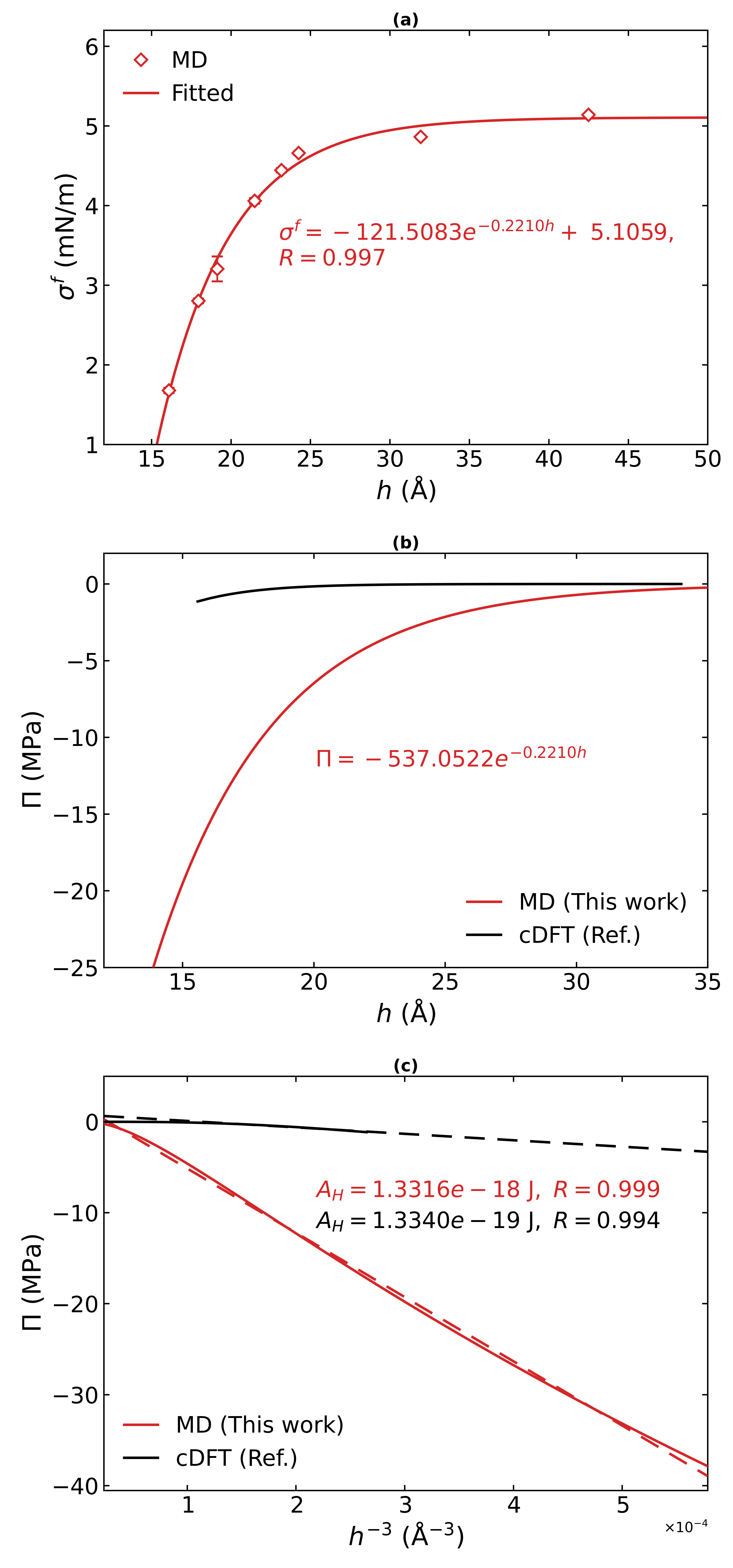}
		\caption{Interfacial properties of gas nanofilms at 120 K with a surface area of $49.056 \times 49.056$ $\mathrm{\AA}^2$: (a) film surface tension, $\sigma^f$, as a function of film thickness, $h$; (b) disjoining pressure, $\Pi$, as a function of film thickness, $h$; and (c) $\Pi$ as a function of $h^{-3}$. Dashed lines represent linear fits. The cDFT results shown in (b) and (c) are reproduced from Ref.~\cite{yang2026accounting}.
		}
		\label{fig:z2}
	\end{centering}
\end{figure}

Fig. \ref{fig:z2}a shows the dependence of film surface tension on film thickness for gas nanofilms at 120 K, with a surface area of $49.056 \times 49.056$ $\mathrm{\AA}^2$, as obtained from MD simulations. The surface tension increases monotonically with increasing film thickness over the range of 15-50 $\mathrm{\AA}$. 
This increasing trend exhibits an exponential dependence~\cite{yang2026resolving,li2024hygroscopic,jianzhou2026estimating}, as indicated by the solid line fitted to the MD data. At large film thicknesses, the surface tension converges to the bulk value of approximately 5.11 mN/m, which is in excellent agreement with the experimental value of 5.06 mN/m~\cite{linstrom2001nist} and the simulation data reported previously~\cite{mendez2022argon}. 

Fig. \ref{fig:z2}b shows the corresponding disjoining pressure isotherm calculated from Eq.~\eqref{eq:8882}. The disjoining pressures are consistently negative, indicating that the bulk gas phase has a higher pressure than the bulk liquid phase. Consequently, the curvature of the meniscus in the Plateau border is opposite to that depicted in Fig.~\ref{fig:z1}a, which is provided solely as a schematic illustration. Notably, negative disjoining pressures have also been reported for liquid nanofilms of pure argon~\cite{bhatt2002molecular,yang2026resolving,jianzhou2026estimating}.

According to classical Hamaker theory, the disjoining pressure can be expressed as~\cite{israelachvili2011intermolecular}:
\begin{equation}
	\label{eq:10}
	\Pi = -\frac{A_H}{6\pi h^3},
\end{equation}
where $A_H$ represents the Hamaker constant. Fig. \ref{fig:z2}c presents the disjoining pressure as a function of $1/h^3$, together with linear fits based on Eq.~\eqref{eq:10}. The MD-predicted disjoining pressures agree well with the theoretical relationship. The Hamaker constant obtained from the linear fit is $1.3316 \times 10^{-18}$ J.

\subsubsection{Film stability}
At small film thicknesses, the film surface tension decreases rapidly, while the magnitude of the disjoining pressure increases exponentially, promoting the destabilization and eventual rupture of the planar gas film. To elucidate the morphological evolution of the gas film, the time-dependent gas-phase configurations were analyzed. 

\begin{figure}[tb]
	\begin{centering}
		\includegraphics[width=0.6\textwidth]{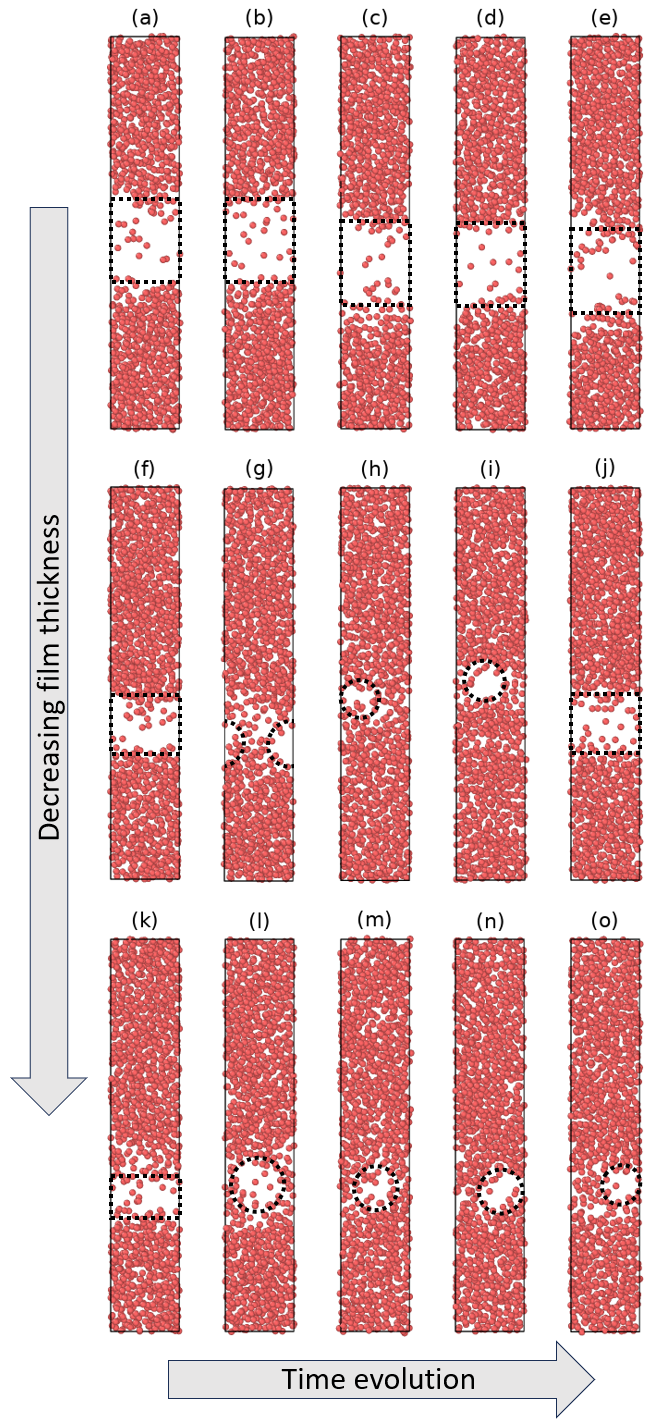}
		\caption{Time evolution of gas-phase morphologies for systems with a surface area of $49.056 \times 49.056$ $\mathrm{\AA}^2$ at 120 K and different initial planar film thicknesses. Planar gas films are enclosed by black dotted boxes, while spherical gas bubbles are indicated by black dotted circles.
		}
		\label{fig:z3}
	\end{centering}
\end{figure}

Three distinct types of morphological evolution were identified, with representative snapshots shown in Fig.~\ref{fig:z3}.
At relatively large film thicknesses (Fig.~\ref{fig:z3}a-e), the gas film retains its planar morphology throughout the simulation, indicating that the planar configuration remains stable over the investigated time scale. At moderate film thicknesses (Fig.~\ref{fig:z3}f-j), the initially planar gas film undergoes a transient morphological transition. A liquid bridge first forms across the gas film and subsequently expands radially while maintaining an approximately cylindrical shape. As the bridge grows, the initially planar gas film progressively transforms into a spherical gas bubble. The spherical bubble, however, persists only for a short period before disappearing, after which the planar gas film is restored.
At extremely small film thicknesses (Fig.~\ref{fig:z3}k-o), a similar morphological transition is observed. In this case, however, the resulting spherical gas bubble persists for the remainder of the simulation, and the planar gas film is not recovered within the investigated time scale. These observations demonstrate that decreasing the film thickness progressively destabilizes the planar gas configuration and promotes the transition toward a spherical gas morphology.

The observed morphological transitions can be qualitatively interpreted in terms of surface free-energy minimization. In a simplified description, the surface free energy can be expressed as~\cite{firoozabadi2016thermodynamics}:
\begin{equation}
	G = \gamma A,
\end{equation}
where $\gamma$ is the surface tension and $A$ is the total surface area. Thus, the stable morphology tends to minimize the total surface free energy, which is determined by the combined effects of the surface tension and surface area.

For large film thicknesses, a hypothetical perturbation leading to the formation of a liquid bridge would substantially increase the total interfacial area. Such an increase in interfacial area would raise the total interfacial free energy, and the planar gas film therefore remains the energetically favorable configuration. This is consistent with the absence of any morphological transition during the simulation. At intermediate film thicknesses, the interfacial areas of the planar and nonplanar configurations are relatively comparable, suggesting that the difference in their total interfacial free energies is relatively small. Consequently, a transient transition from the planar gas film to a spherical gas bubble can occur, although the planar configuration is eventually recovered. In contrast, at extremely small film thicknesses, the formation of a spherical gas bubble substantially reduces the total interfacial area. Combined with the reduced film surface tension at small film thicknesses, this reduction in interfacial area results in a lower total interfacial free energy. Consequently, the spherical gas bubble persists throughout the remainder of the simulation, and the planar gas film is not recovered within the investigated time scale.

It is worth noting that the stability and rupture of liquid nanofilms have been investigated in previous studies~\cite{filippini2014communication,peng2016quantitative}. Similar thickness-dependent rupture behavior has been reported for liquid nanofilms, although the morphological evolution differs from that observed here for gas nanofilms. In liquid nanofilms, rupture is typically initiated by the formation and growth of localized holes as the film thickness decreases. In contrast, the present gas nanofilms undergo a distinct morphological pathway, in which a liquid bridge forms across the planar gas film and subsequently develops into a spherical gas bubble. 

Moreover, Filippini \et~\cite{filippini2014communication} argued that the surface tension of liquid nanofilms remains essentially independent of film thickness as long as the film remains intact, attributing the apparent decrease in surface tension primarily to the presence of holes rather than to confinement effects. In contrast, Peng \et~\cite{peng2016quantitative} identified a two-stage thickness-dependent behavior of aqueous nanofilms, with distinct stability regimes supported by the corresponding disjoining-pressure profiles. In the present study, surface tension was evaluated only for intact planar gas films, excluding configurations after rupture. The observed systematic decrease in surface tension with decreasing film thickness therefore cannot be attributed to ruptured configurations. Taken together, these results provide evidence for genuine thickness-dependent surface properties in nanofilms, consistent with the regime-dependent behavior proposed by Peng \et~\cite{peng2016quantitative}.

\subsubsection{Comparison with cDFT}

We further compare the MD results with corresponding one-dimensional cDFT predictions obtained using the perturbed-chain statistical associating fluid theory (PC-SAFT) equation of state from Yang \et~\cite{yang2026accounting}. The comparisons are presented in Figs.~\ref{fig:z2}b and c. A substantial discrepancy is observed between the two approaches, with cDFT predicting considerably larger disjoining pressures than those obtained from MD simulations. This discrepancy is also reflected in the estimated Hamaker constants, which differ by approximately one order of magnitude. Similar discrepancies between MD simulations and cDFT predictions have also been reported for liquid nanofilms of pure argon~\cite{bhatt2002molecular,yang2026accounting}. Two explanations have been proposed for this discrepancy. Bhatt \textit{et al.}~\cite{bhatt2002molecular} attributed it to limitations of the underlying density functional. However, consistent with the present observations, our previous study~\cite{yang2026accounting} demonstrated that the discrepancy persists even when an advanced PC-SAFT-based functional is employed, suggesting that deficiencies in the underlying density functional alone cannot fully account for the observed discrepancy

An alternative interpretation involves capillary-wave fluctuations of the gas-liquid surfaces~\cite{yang2026accounting}. At finite temperatures, thermal motion of the molecules continuously induces nanoscale fluctuations of the interfacial position, resulting in local variations in the film thickness~\cite{benet2014disjoining,macdowell2014disjoining}. These fluctuations become particularly important for thin films, where even relatively small interfacial displacements can produce substantial changes in the local film thickness and thereby affect the stability and thermodynamic properties of the film. Such fluctuations are naturally captured in MD simulations through the instantaneous molecular configurations. In contrast, conventional cDFT is based on a mean-field description in which the equilibrium density is laterally averaged, and the instantaneous interfacial fluctuations are not explicitly resolved~\cite{bhatt2002molecular,macdowell2014disjoining}. Consequently, capillary-wave fluctuations may contribute to the difference between the disjoining pressures predicted by MD and cDFT. Although this provides a plausible explanation for the observed discrepancy, its role has not yet been directly verified.

To further examine this hypothesis, the following section investigates the effect of film surface area, which is expected to influence the magnitude of capillary-wave fluctuations~\cite{bhatt2002molecular}.

\subsection{Effect of Film Surface Area}

\subsubsection{Surface property isotherms}
\begin{figure}[tb]
	\begin{centering}
		\includegraphics[width=0.5\textwidth]{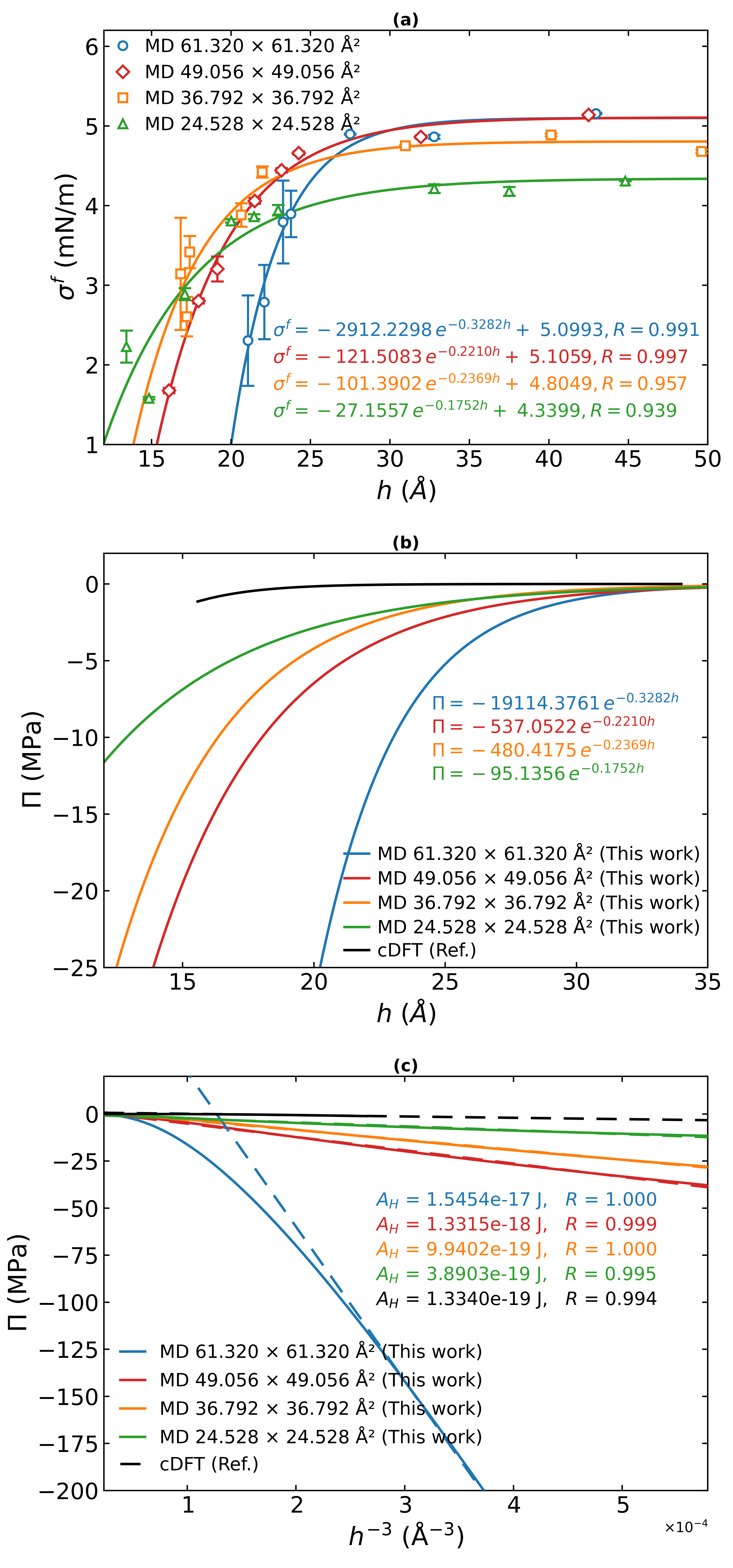}
		\caption{Interfacial properties of gas nanofilms at 120 K with different surface areas: (a) film surface tension, $\sigma^f$, as a function of film thickness, $h$; (b) disjoining pressure, $\Pi$, as a function of film thickness, $h$; and (c) $\Pi$ as a function of $h^{-3}$. Dashed lines represent linear fits. The cDFT results shown in (b) and (c) are reproduced from Ref.~\cite{yang2026accounting}.
		}
		\label{fig:z4}
	\end{centering}
\end{figure}

Fig.~\ref{fig:z4}a shows the film surface tension as a function of film thickness for different surface areas at 120~K. The corresponding fitted curves are shown as solid lines. Clearly, the surface area has a pronounced effect on the thickness dependence of the surface tension. As the surface area increases, the variation in surface tension with film thickness becomes substantially more pronounced, which is consistent with the increasing magnitude of prefactor $a$ in Eq.~\eqref{eq:888}. In particular, a 250\% increase in surface area results in an approximately two-order-of-magnitude increase in $a$. For smaller surface areas, the surface tension exhibits relatively large uncertainties at low film thicknesses and deviates from the exponential dependence, resulting in a relatively low Pearson correlation coefficient, $R$.

At large film thicknesses, the surface tension obtained for smaller surface areas converges to lower bulk surface tension values. This behavior can be attributed to the use of a smaller cutoff distance for the short-range interactions, which is determined as half of the shortest dimension of the simulation box. Consequently, reducing the surface area decreases the lateral box dimensions and, hence, the cutoff distance, leading to a systematic reduction in the calculated bulk surface tension.

It is worth noting that a previous study of aqueous liquid nanofilms reported a pronounced dependence of the surface tension on surface area, which gradually diminished and eventually became negligible at a sufficiently large surface area of $40.0 \times 40.0$ $\mathrm{\AA}^2$~\cite{peng2016quantitative}. This observation may be related to the relatively large film thickness considered in that study, for which the surface tension approaches its bulk-limit value. Our results provide a more complete picture of the surface-area effect. At large film thicknesses, the surface tension indeed becomes less sensitive to surface area as the film approaches the bulk limit. In contrast, at small film thicknesses, a pronounced surface-area dependence persists even at relatively large surface areas. This behavior is associated with enhanced capillary-wave fluctuations in thinner films, as discussed below.

Fig.~\ref{fig:z4}b shows the corresponding disjoining pressure isotherms. The magnitude of the disjoining pressure is substantially reduced as the surface area increases. Notably, the MD results obtained for smaller surface areas show better agreement with the cDFT predictions. This observation provides further evidence that capillary-wave fluctuations contribute significantly to the discrepancy between the MD and cDFT results. In particular, reducing the surface area suppresses long-wavelength capillary-wave fluctuations, thereby stabilizing the planar film and yielding behavior closer to the idealized mean-field description of cDFT. Since conventional cDFT does not explicitly account for capillary-wave fluctuations, its predictions correspond to a mean-field description of an idealized, fluctuation-free interface. In contrast, MD simulations with finite surface areas inherently capture capillary-wave fluctuations, with long-wavelength modes progressively suppressed as the surface area decreases.

Fig.~\ref{fig:z4}c presents the disjoining pressure as a function of $1/h^3$, together with linear fits based on Eq.~\eqref{eq:10}. For relatively small surface areas, the disjoining pressure exhibits an approximately linear dependence on $1/h^3$ over a broad range of film thicknesses, indicating that the Hamaker theory provides a reasonable description of the thickness dependence. However, for the largest surface area, a pronounced deviation from linearity emerges at larger film thicknesses. This deviation indicates that the classical $1/h^3$ Hamaker scaling does not fully describe the MD results over the entire thickness range considered. Because the Hamaker expression is derived under simplified continuum and slab-profile assumptions, deviations can arise when molecular-scale structure and fluctuation effects become important\cite{bhatt2002molecular}. Similar deviations from the classical Hamaker scaling have also been reported in previous molecular simulation studies of liquid nanofilms~\cite{bhatt2002molecular,yang2026resolving}. 

The corresponding Hamaker constants are also indicated in Fig.~\ref{fig:z4}c. As expected, the discrepancy between the Hamaker constants obtained from MD and cDFT decreases systematically with decreasing surface area. Specifically, the difference decreases from approximately two orders of magnitude at the largest surface area to only a factor of approximately three at the smallest surface area. This further supports the conclusion that the discrepancy between MD and cDFT predictions is strongly influenced by capillary-wave fluctuations, which become increasingly suppressed as the surface area decreases.

\subsubsection{Density distributions}

\begin{figure}[tb]
	\begin{centering}
		\includegraphics[width=0.9\textwidth]{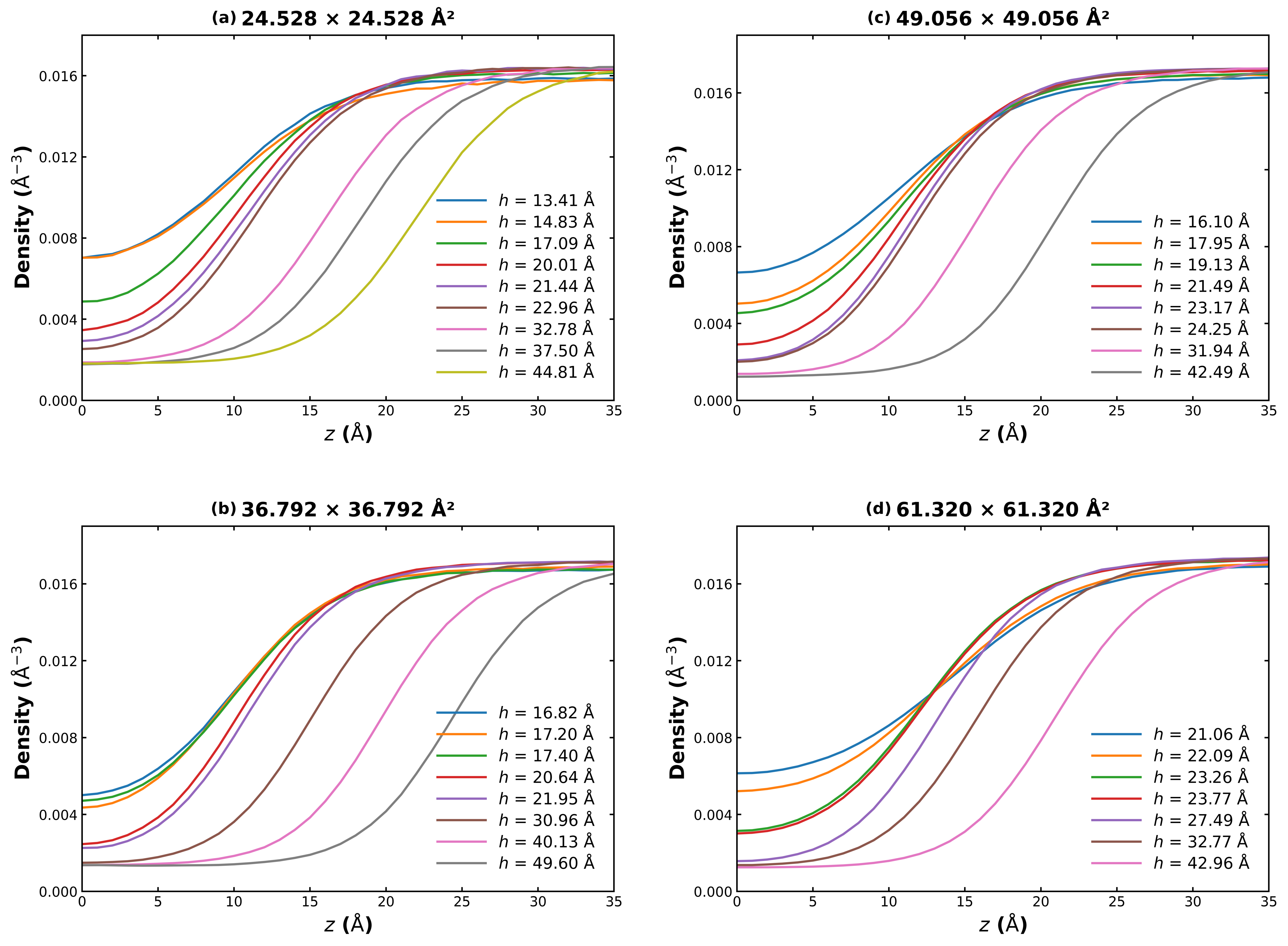}
		\caption{Density distributions from the center of the gas film to the bulk liquid regions for different surface areas at 120 K: (a) $24.528 \times 24.528$ $\mathrm{\AA}^2$; (b) $36.792 \times 36.792$ $\mathrm{\AA}^2$; (c) $49.056 \times 49.056$ $\mathrm{\AA}^2$; and (d) $61.320 \times 61.320$ $\mathrm{\AA}^2$.
		}
		\label{fig:z5}
	\end{centering}
\end{figure}

The density profiles extracted from the center of the nanofilm to the bulk liquid region are presented in Fig. \ref{fig:z5} for various surface areas at 120 K. The interfacial structure of the LJ fluid exhibits a characteristic sigmoidal (tanh-like) shape. A notable observation is the significant increase in the center-film density with decreasing film thickness, which is consistent with the enhanced contribution of attractive interactions (negative disjoining pressure) within the confined liquid. Crucially, the slope of the density transition from the film center to the bulk liquid becomes progressively less steep as the film thickness decreases, indicating a marked softening of the intrinsic density profile. This softening is further evidenced by the crossing of density profiles for films of different thicknesses, with the intersection point shifting to smaller thicknesses for larger surface areas. This behavior is fundamentally distinct from the predictions of cDFT, which, as previously reported, yields density profiles with essentially constant slopes regardless of film thickness~\cite{yang2026accounting}. The observed softening of the density profiles and the associated crossing behavior are direct manifestations of thermally excited capillary-wave fluctuations. These fluctuations increasingly blur the interface as the film becomes thinner and as the surface area increases, resulting in a reduction of the effective density gradient, a signature that is inherently absent in the mean-field description of cDFT.

\subsection{Effect of Temperature}

\subsubsection{Surface property isotherms}
\begin{figure}[tb]
	\begin{centering}
		\includegraphics[width=0.5\textwidth]{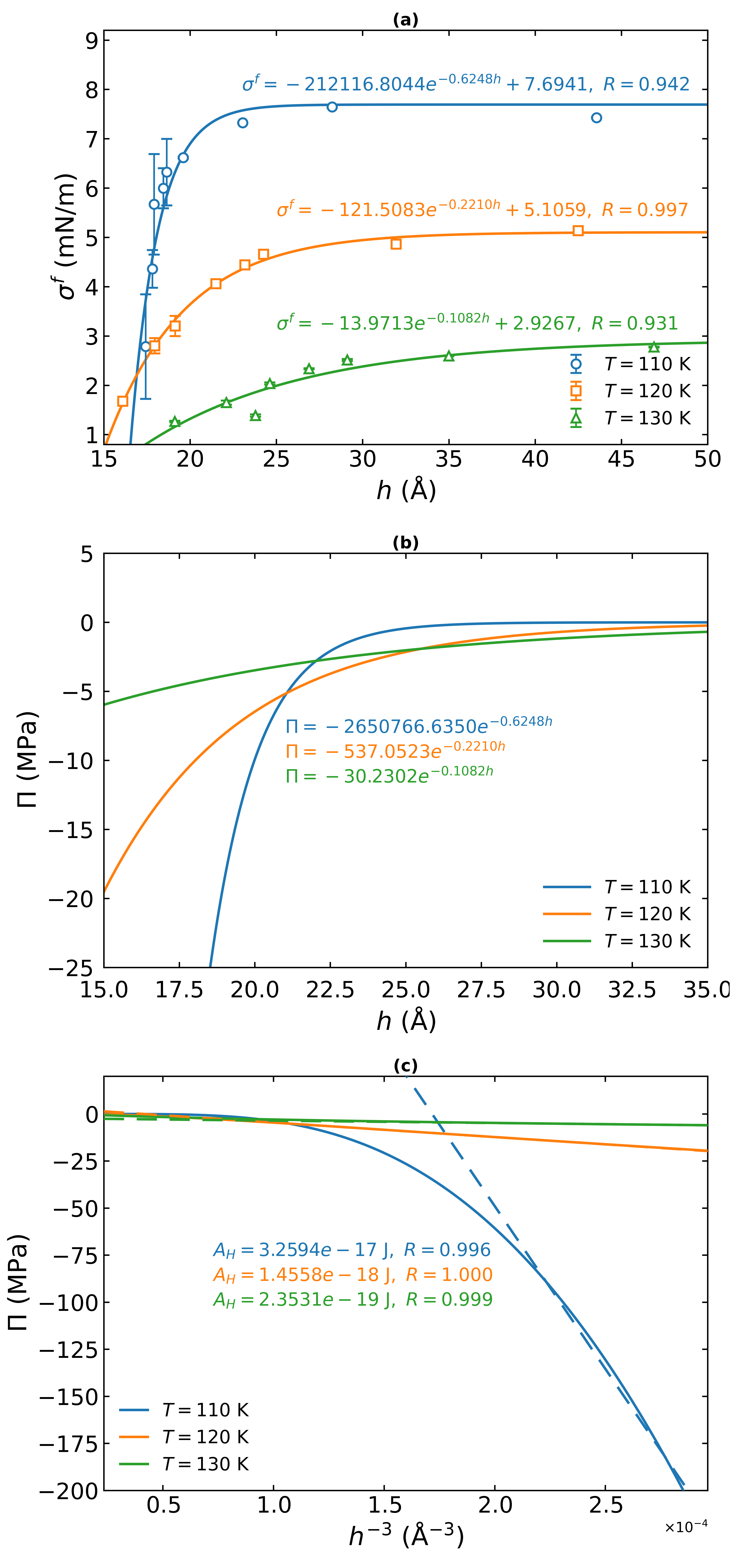}
		\caption{Interfacial properties of gas nanofilms at different temperatures with a surface area of $49.056 \times 49.056$ $\mathrm{\AA}^2$: (a) film surface tension, $\sigma^f$, as a function of film thickness, $h$; (b) disjoining pressure, $\Pi$, as a function of film thickness, $h$; and (c) $\Pi$ as a function of $h^{-3}$. Dashed lines represent linear fits.
		}
		\label{fig:z6}
	\end{centering}
\end{figure}

Fig. \ref{fig:z6}a shows the dependence of the film surface tension on film thickness at a fixed lateral size of $49.056 \times 49.056$ $\mathrm{\AA}^2$ and temperatures of 110, 120, and 130 K. The surface tensions under bulk conditions at 110 and 130 K are 7.69 and 2.92 mN/m, respectively. These values are in good agreement with the experimental data reported by NIST~\cite{linstrom2001nist} and the simulation results available in the literature~\cite{mendez2022argon}. The temperature has a pronounced effect on the thickness dependence of the film surface tension. At higher temperatures, the surface tension varies more gradually with film thickness and approaches its bulk value only at relatively large film thicknesses. In contrast, at lower temperatures, the surface tension changes more rapidly and reaches its bulk value at a smaller film thickness.

This behaviour originates from the temperature-dependent sharpness of the gas-liquid surface, which governs how effectively the two interfaces interact across the film. Interfacial coupling is mediated by short-range van der Waals forces and requires spatial overlap of the density transition layers of the two surfaces. At low temperature, the interface is intrinsically sharp, so the overlap region is extremely narrow and the interfacial coupling is strong; as the film thickens, this overlap is lost abruptly, causing a sudden decoupling of the two surfaces and a rapid recovery of bulk-like properties. Consequently, the surface tension reaches its bulk limit over a short distance. At high temperature, thermal broadening renders the interface diffuse and the transition layers much wider. The overlap persists over a larger range of film thicknesses, and the decay of coupling is correspondingly gradual. Thus, the two surfaces remain partially correlated over a longer scale, leading to a slow, asymptotic convergence of the surface tension to the bulk value. The overall picture highlights a competition between interfacial sharpness and the finite range of intermolecular forces, with the former strongly modulated by temperature.

Fig. \ref{fig:z6}b presents the disjoining pressure isotherms, revealing a complex dependence on temperature. At relatively small film thicknesses, higher temperatures correspond to smaller magnitudes of the disjoining pressure. In contrast, this trend is reversed at larger film thicknesses, where the magnitude of the disjoining pressure increases with temperature. At intermediate film thicknesses, the disjoining pressure exhibits a nonmonotonic temperature dependence, with the isotherms showing a crossover behavior. The corresponding fitted Hamaker constants are shown in Fig. \ref{fig:z6}c. A pronounced temperature dependence is observed, with the fitted Hamaker constant increasing substantially as the temperature decreases. In particular, reducing the temperature from 130 to 110 K results in an approximately 140-fold increase in the fitted Hamaker constant.

The temperature dependence observed here differs markedly from that reported for liquid nanofilms~\cite{bhatt2002molecular,peng2015methodology,yang2026resolving,jianzhou2026estimating,peng2016quantitative}. For liquid nanofilms, higher temperatures generally lead to lower disjoining pressures at relatively small film thicknesses, whereas the opposite trend is observed for the gas nanofilms considered here. This contrasting behavior originates from the fundamentally different mechanical responses of the confined phase to thickness variations, which are dictated by their respective compressibilities.

In a liquid nanofilm, the confined phase is dense and relatively incompressible, especially at lower temperatures. When the film thickness is varied, the liquid structure and density remain largely unaltered because the molecules are already tightly packed~\cite{yang2026accounting,werth2013influence}. Consequently, changes in thickness produce only minimal deviations in the thermodynamic state of the confined liquid from its bulk reference, leading to weak thickness-dependent surface forces and thus small disjoining pressures. In contrast, the present system features a highly compressible gas film confined between bulk liquid phases. Because the gas density is intrinsically low and strongly responsive to spatial confinement, any reduction in film thickness forces the gas molecules into a significantly denser, more ordered state. This enhanced sensitivity is particularly pronounced at lower temperatures, where thermal motion is suppressed, and the gas film is more readily compressed. As a result, thickness variations induce substantial changes in the film's thermodynamic state, generating much stronger surface forces and consequently larger disjoining pressure magnitudes at low temperatures. The opposite responses of liquid and gas nanofilms therefore arise primarily from the distinct compressibility of the confined phase: an incompressible liquid resists thickness-induced structural changes, whereas a compressible gas amplifies them.

\subsubsection{Density distributions}

\begin{figure}[tb]
	\begin{centering}
		\includegraphics[width=0.6\textwidth]{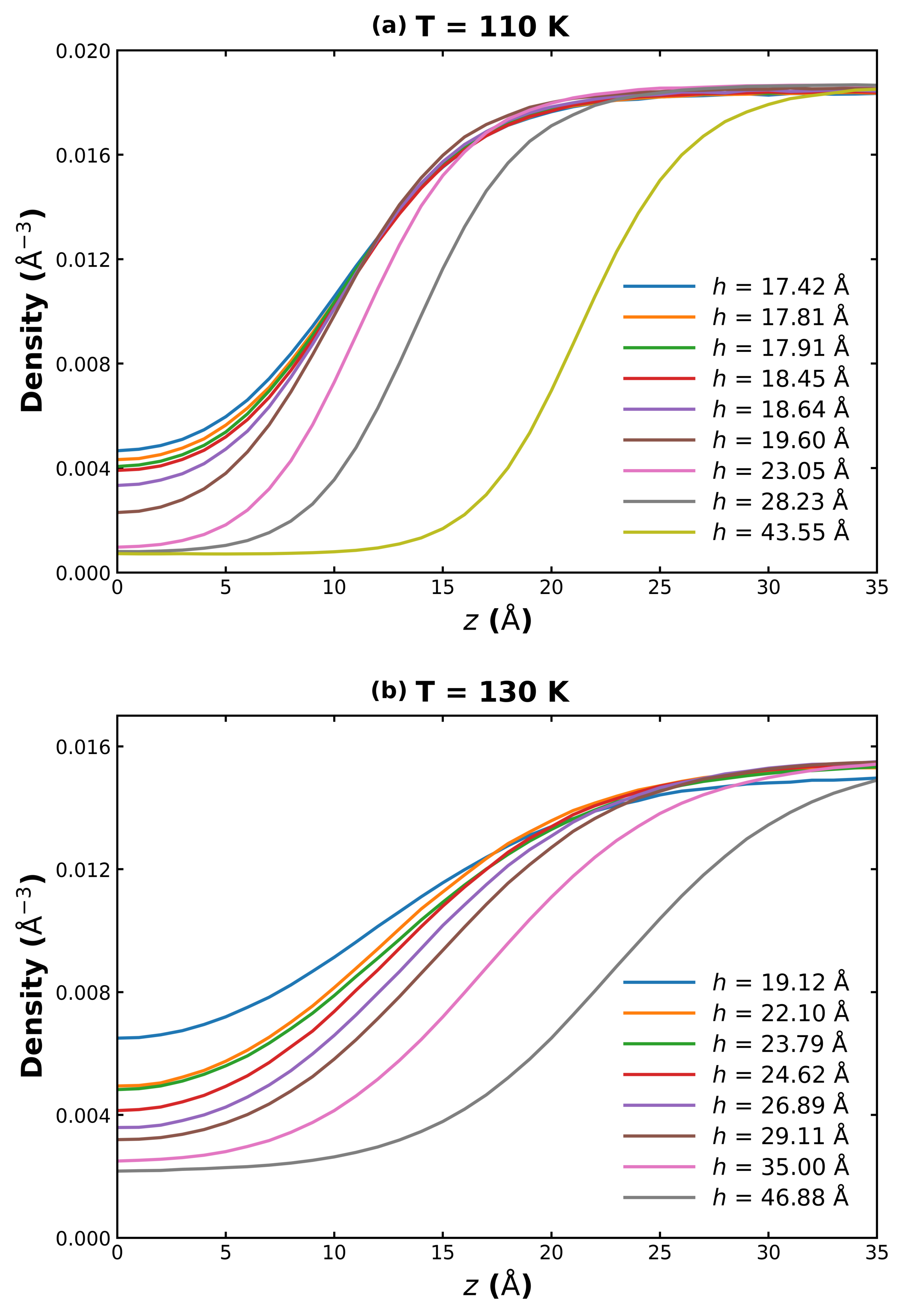}
		\caption{Density distributions from the center of the gas film to the bulk liquid regions for a surface area of $49.056 \times 49.056$ $\mathrm{\AA}^2$ at different temperatures: (a) 110 K and (b) 130 K. 
		}
		\label{fig:z7}
	\end{centering}
\end{figure}

We further examine the density profiles at 110 K and 130 K, shown in Fig. \ref{fig:z7}. For both temperatures, the density distributions exhibit the characteristic sigmoidal transition from the film center to the bulk liquid phase, consistent with the profiles observed at 120 K. A marked temperature dependence is observed: at 110 K, the density transition is notably steeper than at 130 K for comparable film thicknesses. This behavior provides direct structural evidence for the temperature-dependent interfacial sharpness discussed above. At lower temperatures, the gas-liquid surface is intrinsically sharp, resulting in a narrow density transition layer. Consequently, the two interfaces decouple abruptly upon film thickening, leading to a rapid recovery of bulk-like properties and a pronounced sensitivity of the disjoining pressure to thickness variations. In contrast, at higher temperatures, thermal fluctuations broaden the interface, producing a diffuse and extended transition layer. This softening of the density profile enhances the spatial overlap between the two interfaces over a larger range of film thicknesses, giving rise to a gradual, asymptotic convergence of surface properties to their bulk limits.

\section{Conclusion}

In this work, molecular dynamics simulations were employed to systematically investigate the thermodynamics and stability of gas nanofilms confined between bulk liquid phases. A Gibbs surface thermodynamic framework was established to consistently relate the gas film surface tension to the disjoining pressure. Unlike previous studies, which have focused predominantly on liquid nanofilms, the present study provides a direct molecular-level characterization of free gas nanofilms and their stability.

The results demonstrate that the surface properties of gas nanofilms depend strongly on film thickness, surface area, and temperature. At sufficiently large thicknesses, the film surface tension approaches the bulk liquid-gas surface tension, whereas decreasing the film thickness leads to a rapid reduction in surface tension and an increasingly negative disjoining pressure. The latter approximately follows the classical $h^{-3}$ dependence over a substantial thickness range, although deviations from Hamaker scaling emerge at larger film thicknesses. More importantly, the thinning-induced destabilization of gas nanofilms proceeds through a distinct morphological pathway. The gas film undergoes the formation and growth of a liquid bridge, followed by a  transformation into a spherical gas bubble. The transition from a stable planar film to a transient or persistent bubble state reveals a thickness-dependent stability landscape governed by the competition between surface free energy and interfacial area.

The pronounced surface-area dependence provides important insight into the role of thermal fluctuations at the nanoscale. Increasing the lateral area substantially enhances the thickness dependence of the surface tension and the magnitude of the disjoining pressure. In contrast, reducing the surface area suppresses long-wavelength capillary-wave fluctuations and progressively brings the MD results into closer agreement with PC-SAFT cDFT predictions. 
This systematic trend strongly supports the view that the long-standing discrepancy between MD and cDFT predictions of disjoining pressure for free nanofilms is primarily associated with capillary-wave fluctuations, which are explicitly captured in MD simulations but absent from conventional mean-field cDFT. 
These findings highlight the importance of accounting for thermal fluctuations when developing predictive theoretical descriptions of nanoscale free films and help clarify the origin of the discrepancy between molecular simulations and mean-field theories.

Temperature also plays a distinct role in determining the thermodynamic behavior of gas nanofilms. In contrast to the generally reported behavior of liquid nanofilms, the magnitude of the disjoining pressure at small film thicknesses increases with decreasing temperature. This behavior is consistent with the high compressibility of the confined gas nanofilm, which leads to enhanced sensitivity to thickness variations as the temperature decreases. As a result, changes in film thickness produce stronger variations in the thermodynamic state of the gas film, leading to enhanced disjoining pressures under stronger confinement. These results highlight that the temperature dependence of nanofilm thermodynamics is closely coupled to the compressibility of the confined phase; accordingly, this temperature-dependent behaviour differs fundamentally between gas and liquid nanofilms.

Overall, this study extends the molecular understanding of nanofilm thermodynamics from predominantly liquid systems to gas films and demonstrates the coupled effects of film thickness, surface area, temperature, and morphological stability, together with the important influence of thermal fluctuations.
By explicitly resolving molecular fluctuations, this work addresses the limited molecular-level understanding of free gas nanofilms and reveals their distinct stability and thermodynamic behavior.
More broadly, the results highlight that nanofilms are thermodynamically active structures whose properties are strongly influenced by nanoscale confinement and molecular fluctuations. This insight is particularly relevant to confined multiphase processes in nanoporous media, where gas films can influence phase behavior, capillary effects, and fluid transport. The findings therefore provide a molecular basis for understanding gas-film stability and coalescence and for developing more accurate models of nanoscale multiphase phenomena.


\bigskip
{\bf{Acknowledgments\\[1ex]}}
The research is supported by the Fundamental Research Funds for the Central Universities (2025QN1175).


\comment{
	\bigskip
	{\bf{Data and Software Availability\\[1ex]}}
	All data and simulation scripts are available from the corresponding author upon request.
}
\bibliography{Pdis3}

\clearpage

\begin{figure}[tb]
	\centering
	\includegraphics[width=0.6\textwidth]{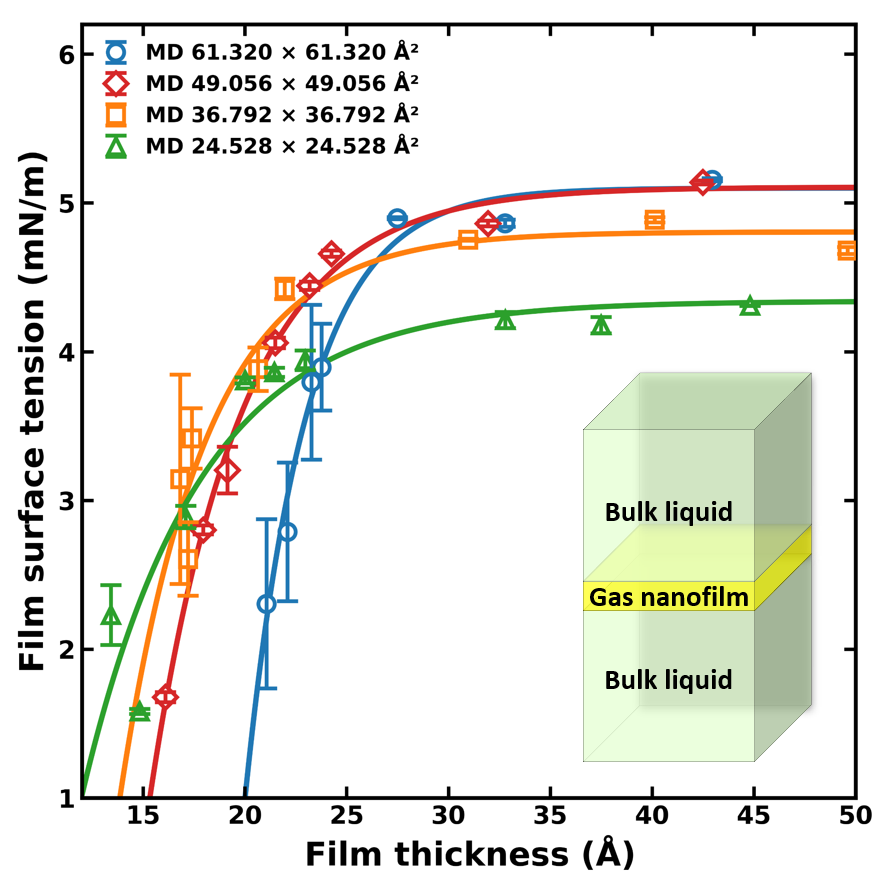}
	
	\vspace{0.5cm}
	
	{\large\bfseries Graphical Abstract\par}
	
	\vspace{1.5ex}
	
	\noindent\parbox{\textwidth}{%
		\setlength{\parindent}{2em}%
		\indent Dependence of surface tension on film thickness for planar gas nanofilms with different surface areas confined between two bulk liquid phases of Lennard-Jones argon at 120 K.
	}
\end{figure}

\clearpage

\clearpage

\clearpage

\clearpage

\clearpage

\clearpage

\clearpage

\end{document}